\documentclass[epj]{webofc}
\usepackage[utf8]{inputenc}
\usepackage[varg]{txfonts}   % Web of Conferences font
\usepackage{booktabs}
\usepackage{xcolor}
\definecolor{darkred}{rgb}{0.4,0.0,0.0}
\definecolor{darkgreen}{rgb}{0.0,0.4,0.0}
\definecolor{darkblue}{rgb}{0.0,0.0,0.4}
\usepackage[bookmarks,linktocpage,colorlinks,
    linkcolor = darkred,
    urlcolor  = darkblue,
    citecolor = darkgreen]{hyperref}
\usepackage{subfigure}
\wocname{EPJ Web of Conferences}
\woctitle{Lattice2017}
\usepackage{graphicx}
\begin{document}
%%%%%%%%%%%%%%%%%%%%%%%%%%%%%%%%%%%%%%%%%%%%%%%%%%%%%%%%%%%%%%%%%%%%%%%%%%%%%
%
\selectlanguage{english}
%----------------------------------------------------------------------------
\title{%
Baryon interactions from lattice QCD with physical quark masses
-- Nuclear forces and $\Xi\Xi$ forces --
}
%----------------------------------------------------------------------------
\author{%
  \firstname{Takumi} \lastname{Doi}\inst{1,2}\fnsep\thanks{%
    Speaker, \email{doi@ribf.riken.jp}} \and
%    Acknowledges financial support by his mentor J.R.R. Tolkien.} \and
  \firstname{Takumi}    \lastname{Iritani}\inst{1} \and
  \firstname{Sinya}     \lastname{Aoki}\inst{1,3,4} \and
  \firstname{Shinya}    \lastname{Gongyo}\inst{1} \and
  \firstname{Tetsuo}    \lastname{Hatsuda}\inst{1,2} \and
  \firstname{Yoichi}    \lastname{Ikeda}\inst{1,5} \and
  \firstname{Takashi}   \lastname{Inoue}\inst{1,6} \and
  \firstname{Noriyoshi} \lastname{Ishii}\inst{1,5} \and
  \firstname{Takaya}    \lastname{Miyamoto}\inst{1,3} \and
%  \firstname{Keiko}     \lastname{Murano}\inst{1,5} \and
  \firstname{Hidekatsu} \lastname{Nemura}\inst{1,5} \and
  \firstname{Kenji}     \lastname{Sasaki}\inst{1,3}
}
%----------------------------------------------------------------------------
\institute{%
  Theoretical Research Division, Nishina Center, RIKEN, Wako 351-0198, Japan
  \and
  iTHEMS Program and iTHES Research Group, RIKEN, Wako 351-0198, Japan
  \and
  Center for Gravitational Physics, Yukawa Institute for Theoretical Physics, Kyoto University, Kyoto 606-8502, Japan
  \and
  Center for Computational Sciences, University of Tsukuba, Tsukuba 305-8577, Japan
  \and
  Research Center for Nuclear Physics (RCNP), Osaka University, Osaka 567-0047, Japan
  \and
  Nihon University, College of Bioresource Sciences, Kanagawa 252-0880, Japan
}
%----------------------------------------------------------------------------
\abstract{%
  We present the latest lattice QCD results for baryon interactions
  obtained at nearly physical quark masses.
  $N_f = 2+1$ nonperturbatively ${\cal O}(a)$-improved Wilson quark action with stout smearing
  and Iwasaki gauge action are employed on the lattice of
  $(96a)^4 \simeq (8.1\mbox{fm})^4$ with $a^{-1} \simeq 2.3$ GeV,
  where $m_\pi \simeq 146$ MeV and $m_K \simeq 525$ MeV.
  In this report, we study the two-nucleon systems and two-$\Xi$ systems
  in $^1S_0$ channel and $^3S_1$-$^3D_1$ coupled channel,
  and extract central and tensor interactions by the HAL QCD method.
  We also present the results for the $N\Omega$ interaction in $^5S_2$ channel
  which is relevant to the $N\Omega$ pair-momentum correlation in heavy-ion collision experiments.
}
%----------------------------------------------------------------------------
\maketitle
%----------------------------------------------------------------------------
%\vspace*{-17mm}
%\vspace*{-5mm}
\section{Introduction}
%\vspace*{-1mm}
\label{sec:intro}

In quest of the coherent unification of
physics in different hierarchies,
the baryon interactions serve as
``the bridge from quarks to nuclei and cosmos''
(particle-, nuclear- and astro- physics)~\cite{Doi:2017niz}.
Nuclear forces and hyperon forces govern
the properties of (hyper) nuclei, and have been the subject of 
intensive theoretical/experimental investigations.
These interactions
%are also expected to
also
play a crucial role in the equation of state (EoS)
of high dense matter, which is realized at the core of neutron stars.
%and/or in the center of supernova explosions.
%
%
The recent observation of the binary neutron star merger~\cite{TheLIGOScientific:2017qsa,GBM:2017lvd},
%from the gravitational waves and electromagnetic waves
which opens the new era of multi-messenger astronomy,
certainly adds new urgency to the determination of baryon interactions.
%In order to listen and understand these ``messages'',
%the detailed information on baryon interactions plays a crucial role.

%To date,
%the precision interactions have been established
%for two-nucleon ($NN$) forces in phenomenological models and/or
%chiral effective field theories.
%They, however, contain a few tens of parameters to be determined
%from experimental inputs,
%and the connection to the fundamental theory, quantum chromodynamics (QCD),
%is yet to be established.
%Hyperon forces suffer from large uncertainties due to the difficulty of
%scattering experiments with (short-lived) hyperon(s).
%The determination of three-baryon forces, in particular with hyperon(s),
%is much more challenging, although their large impact on EoS of dense matter
%has been well recognized.

In view of the importance of the first-principles calculations of baryon forces,
we
%launched
started
the first realistic lattice QCD (LQCD) calculation
of baryon interactions at nearly physical quark masses about two years ago,
where the interaction kernels (so-called ``potentials'')
are determined by the HAL QCD method~\cite{Ishii:2006ec,Aoki:2012tk}.
New challenges which are inherent in multi-baryon systems on a lattice
are overcome by novel theoretical/algorithmic developments,
most notably by (i) the time-dependent formalism of the HAL QCD method~\cite{HALQCD:2012aa},
(ii) the extension to coupled channels systems~\cite{Aoki:2011gt} and
(iii) the unified contraction algorithm~\cite{Doi:2012xd}.
In particular, time-dependent HAL QCD method enables us to extract 
baryon interactions without relying on the ground state saturation~\cite{HALQCD:2012aa}.
This is the indispensable feature
for a reliable LQCD calculation of baryon interactions,
since a typical excitation energy in multi-baryon systems 
is one to two orders of magnitude smaller than ${\cal O}(\Lambda_{\rm QCD})$ due to
the existence of elastic excited states.
%and the strategy to achieve
%the ground state saturation by taking large Euclidean time $t$ and/or indentification
%of plateau at small $t$ will fail.
This is in contrast to the traditional approach
(so-called ``direct'' calculations~\cite{Yamazaki:2015asa, Orginos:2015aya, Berkowitz:2015eaa}),
which relies on the ground state saturation
and thus is generally
unreliable~\cite{Iritani:2016jie, Iritani:2017rlk, Iritani:2017wvu}.
The existence of uncontrolled systematic errors
in the results of the previous direct calculations
was explicitly exposed by the ``sanity (consistency) check'' proposed in Ref.~\cite{Iritani:2017rlk}.

\begin{figure}[t]
\begin{center}
%
%\vspace*{-4mm}
\includegraphics[angle=0,width=0.85\textwidth]{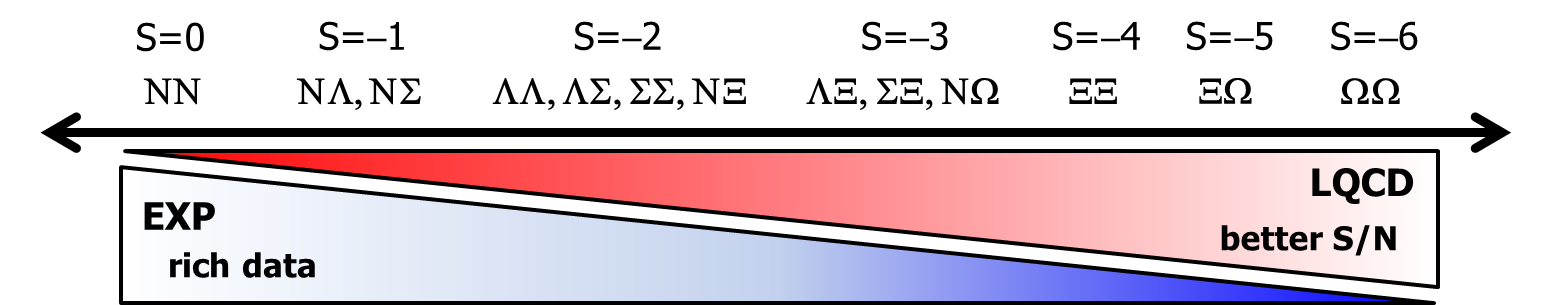}
%\vspace*{-2mm}
%
\caption{
\label{fig:lat_exp}
%
%Schematic picture
Illustrative figure
for the complementary role of lattice QCD and experiments
in the determination of baryon forces.
Lattice QCD has a special strength to
predict interactions in larger strangeness $|S|$ sectors,
where experimental information is scare.
}
\end{center}
%\vspace*{-3mm}
\end{figure}

Our aim in this first physical point calculation %first realistic calculation
is 
the comprehensive determination of
two-baryon interactions from the strangeness $S = 0$ to $-6$
(nuclear forces and hyperon forces)
in parity-even channels. % (S-wave and D-wave).
As is well known, the statistical fluctuations in LQCD
are smaller (larger) for larger (smaller) quark masses,
and thus the results have better precision in larger strangeness $|S|$ sector.
On the other hand, an experiment in larger $|S|$ sector is more difficult
due to the short life time of hyperons.
Therefore, LQCD studies and experiments are complementary with each other
in the determination of baryon forces (See Fig.~\ref{fig:lat_exp}).
Our LQCD prediction on the ``most strange dibaryon'' ($\Omega\Omega$) system
is already available in Ref.~\cite{Gongyo:2017fjb}.

In this paper, we present the latest LQCD results
for two-$\Xi$ ($\Xi\Xi$) forces ($S=-4$) and two-nucleon ($NN$) forces ($S=0$)
in parity-even channel,
updating our previous results~\cite{Doi:2017cfx}.
Central forces are extracted in $^1S_0$ channel,
and central and tensor forces are obtained in $^3S_1$-$^3D_1$ coupled channel analysis.
The results for other interactions between two octet baryons
are presented in Refs.~\cite{Ishii:lat2017}.
Interactions between decuplet and octet baryon also attract great interest and
we have found that the $N\Omega$ system in $^5S_2$ channel
is attractive strong enough to form a bound state at heavy quark masses~\cite{Etminan:2014tya}.
In this report, we present the first physical point results of the $N\Omega$ $(^5S_2)$ interaction,
which can be examined through the $N\Omega$ correlation
in relativistic heavy-ion collision~\cite{Morita:2016auo}.

%%%%%%%%%%%%%%%%%%%%%%%%%%%%%%%%%%%%%%%%%%%%%%%%%%
%%%%%%%%%%%%%%%%%%%%%%%%%%%%%%%%%%%%%%%%%%%%%%%%%%

%%%%%%%%%%%%%%%%%%%%%%%%%%%%%%%%%%%%%%%%%%%%%%%%%%%%
%\vspace*{-1mm}
\section{Formalism}
%\vspace*{-1mm}
\label{sec:formalism}

In the HAL QCD method, 
the key quantity is the equal-time Nambu-Bethe-Salpeter (NBS) wave function.
In the case of the $NN$ system, for instance,
it is defined by
%
%
%\begin{eqnarray}
$
\phi_W^{NN}(\vec{r}) \equiv 
1/Z_N \cdot
\langle 0 | N(\vec{r},0) N(\vec{0},0) | NN, W \rangle_{\rm in} ,
$
%\end{eqnarray}
%
where 
$N$ is the nucleon operator
with its wave-function renormalization factor $\sqrt{Z_N}$ 
and
$|NN, W \rangle_{\rm in}$ denotes the asymptotic in-state of the $NN$ system 
at the total energy of $W = 2\sqrt{k^2+m_N^2}$
with 
%the nucleon mass $m_N$ and 
the asymptotic momentum $k$,
and we consider the elastic region,  $W < W_{\rm th} = 2m_N + m_\pi$.
The most important property of the NBS wave function is that
the information of the phase shift $\delta_l(k)$ ($l$: the orbital angular momentum)
is encoded in the asymptotic behavior
at $r \equiv |\vec{r}| \rightarrow \infty$
as
%
%\begin{eqnarray}
$
\phi_W^{NN} (\vec{r}) \propto 
%\frac{\sin(kr-l\pi/2 + \delta_l^W)}{kr}, 
\sin(kr-l\pi/2 + \delta_l(k)) / (kr).
%\quad 
%r \equiv |\vec{r}| \rightarrow \infty,
$
%\end{eqnarray}
%
%
Exploiting this feature,
one can define non-local but energy-independent %$NN$
potential, $U^{NN}(\vec{r},\vec{r}')$,
which is faithful to the phase shifts
through the Schr\"odinger equation~\cite{Ishii:2006ec, HALQCD:2012aa, Aoki:2012tk},
%
%\begin{eqnarray}
%
$
(E_W^{NN} - H_0) \phi_W^{NN}(\vec{r})
= 
\int d\vec{r'} U^{NN}(\vec{r},\vec{r'}) \phi_W^{NN}(\vec{r'}) ,
$
%\label{eq:Sch_2N:tindep}
%\end{eqnarray}
%
where 
$H_0 = -\nabla^2/(2\mu)$ and
$E_W^{NN} = k^2/(2\mu)$ with the reduced mass $\mu = m_N/2$.
%
%It has been also proven that 
%$U^{NN}(\vec{r},\vec{r'})$ can be constructed
%as to be energy-independent~\cite{Ishii:2006ec,Aoki:2012tk}.

Generally speaking, the NBS wave function 
can be extracted from the %corresponding 
four-point correlator,
$
G^{NN} (\vec{r},t)
\equiv
%\frac{1}{L^3}
\sum_{\vec{R}}
\langle 0 |
          (N(\vec{R}+\vec{r}) N (\vec{R}))(t)\
\overline{(N N)}(t=0)
| 0 \rangle ,
$ 
by isolating the contribution from each energy eigenstate
(most typically by the ground state saturation with $t \rightarrow \infty$).
Such a procedure, however, is practically almost impossible,
due to the existence of nearby elastic scattering states.
In fact, the typical excitation energy is
as small as ${\cal O}(1)-{\cal O}(10)$ MeV,
which is estimated by the empirical binding energies and/or
the discretization in spectrum by the finite volume, $\sim (2\pi/L)^2 / m_N$.
Correspondingly, ground state saturation 
requires 
$t \gtrsim {\cal O}(10)-{\cal O}(100)$ fm, 
which is far beyond reach considering that
signal/noise (S/N) is exponentially degraded in terms of $t$.

This fundamental problem inherent in multi-baryon systems on a lattice
can be overcome 
%Recently, the breakthrough on this issue was achieved 
by the time-dependent HAL QCD method~\cite{HALQCD:2012aa},
in which the signal of $U^{NN}(\vec{r},\vec{r'})$
is extracted even from elastic excited states,
using the energy-independence of $U^{NN}(\vec{r},\vec{r'})$.
More specifically, 
the following ``time-dependent'' Schr\"odinger equation holds
even without the ground state saturation,
%%%%%%%%%%%%%%%%%%%%%%
%\vspace*{-1mm}
%\begin{eqnarray}
\begin{equation}
\left( 
- \frac{\partial}{\partial t} 
+ \frac{1}{4m_N} \frac{\partial^2}{\partial t^2} 
- H_0
\right)
R^{NN}(\vec{r},t) 
=
\int d\vec{r'} U^{NN}(\vec{r},\vec{r'}) R^{NN}(\vec{r'},t) ,
\label{eq:Sch_2N:tdep}
\end{equation}
%\end{eqnarray}
%\vspace*{-1mm}
%%%%%%%%%%%%%%%%%%%%%%
%
where
$R^{NN}(\vec{r},t) \equiv G^{NN} (\vec{r},t) e^{2m_Nt}$.
While it is still necessary to suppress the contaminations from inelastic states,
it can be fulfilled by much easier condition,
$t \gtrsim (W_{\rm th} - W)^{-1} \sim {\cal O}(1)$ fm.
This is in contrast to the direct calculations~\cite{Yamazaki:2015asa, Orginos:2015aya, Berkowitz:2015eaa}
which inevitably rely on the ground state saturation:
They are generally unreliable
due to the fake (mirage) plateau identification~\cite{Iritani:2016jie, Iritani:2017rlk, Iritani:2017wvu},
and explicit evidences of the uncontrolled systematic errors were also exposed
by the ``sanity (consistency) check'' given in~\cite{Iritani:2017rlk}.

%Note that
%the identification of ``plateau-like'' structures in the effective energies of
%%$\sum_{\vec{r}}G^{NN}(\vec{r},t)$ and/or $\sum_{\vec{r}}R^{NN}(\vec{r},t)$,
%$G^{NN}$ and/or $R^{NN}$,
%which has been customarily used in the previous direct calculations~\cite{Yamazaki:2015asa, Orginos:2015aya, Berkowitz:2015eaa},
%is generally unreliable~\cite{Iritani:2016jie, Iritani:2017rlk, Iritani:2017wvu},
%since one cannot distinguish the correct plateau from the fake plateau (called ``mirage'' in Ref.~\cite{Iritani:2016jie}).
%In fact, the ``sanity (consistency) check'' proposed in Ref.~\cite{Iritani:2017rlk}
%exposed the evidence that the results of previous direct calculations
%suffer from uncontrolled errors.

%%%%%%%%%%%%%%%%%%%%%%%%%%%%%%%%%%%%%%%%%%%%%%%%%%%%
\iffalse
\fi
%%%%%%%%%%%%%%%%%%%%%%%%%%%%%%%%%%%%%%%%%%%%%%%%%%%%

%%%%%%%%%%%%%%%%%%%%%%%%%%%%%%%%%%%%%%%%%%%%%%%%%%
\begin{figure}[t]
\begin{minipage}{0.48\textwidth}
\begin{center}
%
%\vspace*{-4mm}
\includegraphics[angle=0,width=0.90\textwidth]{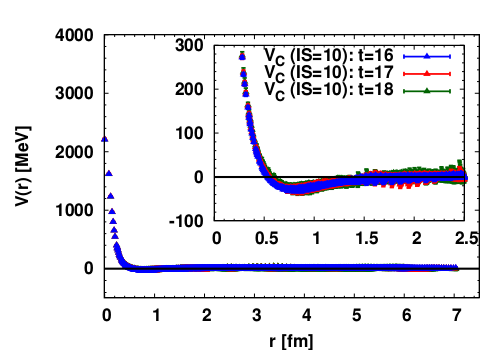}
%
%\vspace*{-2mm}
\caption{
\label{fig:pot:XiXi:1S0:cen}
$\Xi\Xi$ central force $V_C(r)$ in $^1S_0$ $(I=1)$ channel
obtained at $t = 16-18$.
}
\end{center}
\end{minipage}
\hfill
%\vspace*{-2mm}
\begin{minipage}{0.48\textwidth}
\begin{center}
%
%\vspace*{-5mm}
\includegraphics[angle=0,width=0.90\textwidth]{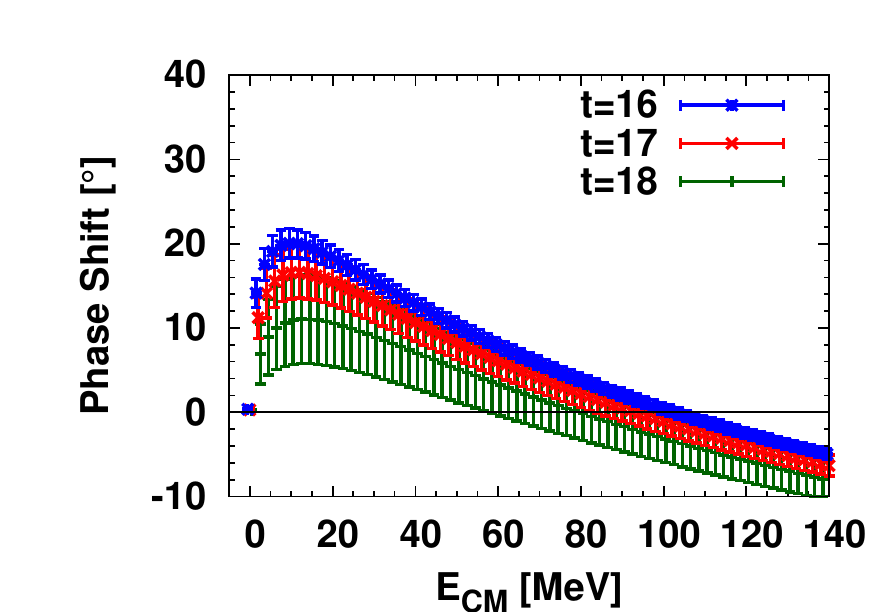}
%
%\vspace*{-2mm}
\caption{
\label{fig:phase:XiXi:1S0:cen}
$\Xi\Xi$ phase shifts in $^1S_0$ $(I=1)$ channel
obtained at $t = 16-18$.
}
\end{center}
\end{minipage}
%\vspace*{-3mm}
\end{figure}
%%%%%%%%%%%%%%%%%%%%%%%%%%%%%%%%%%%%%%%%%%%%%%%%%%

%%%%%%%%%%%%%%%%%%%%%%%%%%%%%%%%%%%%%%%%%%%%%%%%%%%%
%\vspace*{-1mm}
\section{Lattice QCD setup}
%\vspace*{-1mm}
\label{sec:setup}

$N_f = 2+1$ gauge configurations are generated on the $96^4$ lattice
with the Iwasaki gauge action at $\beta = 1.82$ and 
nonperturbatively ${\cal O}(a)$-improved Wilson quark action with $c_{sw} = 1.11$ %~\cite{Taniguchi:2012kk}
and APE stout smearing with $\alpha = 0.1$, $n_{\rm stout} = 6$.
About 2000 trajectories are generated after the thermalization,
and preliminary studies show that $a^{-1} \simeq 2.333$ GeV ($a \simeq 0.0846$ fm)
and $m_\pi \simeq 146$ MeV, $m_K \simeq 525$ MeV. %~\cite{Ishikawa:2015rho}.
The lattice size, $La \simeq 8.1$ fm, is sufficiently large
to accommodate two baryons on a box.
For further details on the gauge configuration generation,
see Ref.~\cite{Ishikawa:2015rho}.

The measurements of NBS correlators are performed at the unitary point,
where the block solver~\cite{Boku:2012zi} is used for the quark propagator
and unified contraction algorithm~\cite{Doi:2012xd} is used for the contraction.
The computation for the measurements (including I/O)
achieves $\sim$ 25\% efficiency, or $\sim$ 65 TFlops sustained on 2048 nodes of K computer.
%For two-octet baryon forces, we calculate all 52 channels relevant in parity-even channel.
We employ wall quark source with Coulomb gauge fixing,
%(which is performed after the stout smearing).
where
the periodic (Dirichlet) boundary condition is used for spacial (temporal) directions
and forward and backward propagations are averaged to reduce the statistical fluctuations.
We pick 1 configuration per each 5 trajectories,
and we make use of the rotation symmetry to increase the statistics.
The total statistics for $NN$ and $\Xi\Xi$ systems
amounts to 414 configurations $\times$ 4 rotations $\times$ 72 wall sources binned by 46 configurations,
and
one for $N\Omega$ system is 
200 configurations $\times$ 4 rotations $\times$ 48 wall sources binned by 10 configurations.

Baryon forces are determined in  $^1S_0$ and $^3S_1$-$^3D_1$ channels
for $NN$ and $\Xi\Xi$ and
in $^5S_2$ channel for $N\Omega$.
We perform the velocity expansion~\cite{Aoki:2012tk} in terms of 
the non-locality of potentials,
and obtain the leading order potentials, i.e., central and tensor forces.
$N\Omega$ $(^5S_2) $ interaction is obtained neglecting the couplings to
two octet baryons (e.g., $\Lambda\Xi$) which are
kinematically suppressed by the D-wave nature.
In this preliminary analysis shown below, 
the term which corresponds to the relativistic effects
($\partial^2 / \partial t^2$-term in Eq.~(\ref{eq:Sch_2N:tdep}))
is neglected.

%%%%%%%%%%%%%%%%%%%%%%%%%%%%%%%%%%%%%%%%%%%%%%%%%%%%
%\vspace*{-1mm}
\section{$\Xi\Xi$ systems ($S=-4$ channel)}
%\vspace*{-1mm}
\label{sec:XiXi}

%%%%%%%%%%%%%%%%%%%%%%%%%%%%%%%%%%%%%%%%%%%%%%%%%%
\begin{figure}[t]
\begin{minipage}{0.48\textwidth}
\begin{center}
%
%\vspace*{-4mm}
\includegraphics[angle=0,width=0.90\textwidth]{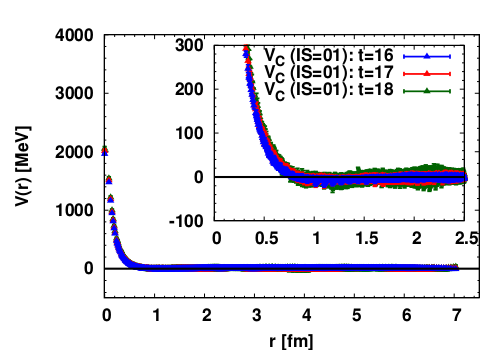}
\end{center}
\end{minipage}
\hfill
%\vspace*{-2mm}
\begin{minipage}{0.48\textwidth}
\begin{center}
%
%\vspace*{-4mm}
\includegraphics[angle=0,width=0.90\textwidth]{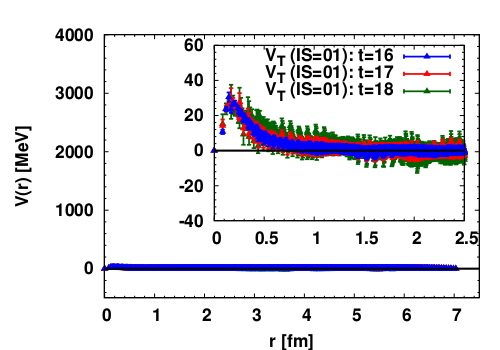}
\end{center}
\end{minipage}
\caption{
\label{fig:pot:XiXi:3S1:cen_ten}
$\Xi\Xi$ central force $V_C(r)$ (left) and tensor force $V_T(r)$ (right) in $^3S_1$-$^3D_1$ $(I=0)$ channel
obtained at $t = 16-18$.
}
%\vspace*{-3mm}
%\end{figure}
%%%%%%%%%%%%%%%%%%%%%%%%%%%%%%%%%%%%%%%%%%%%%%%%%%
%
%
%%%%%%%%%%%%%%%%%%%%%%%%%%%%%%%%%%%%%%%%%%%%%%%%%%
%\begin{figure}[t]
\begin{center}
\sidecaption
%
%\vspace*{-4mm}
\includegraphics[angle=0,width=0.432\textwidth]{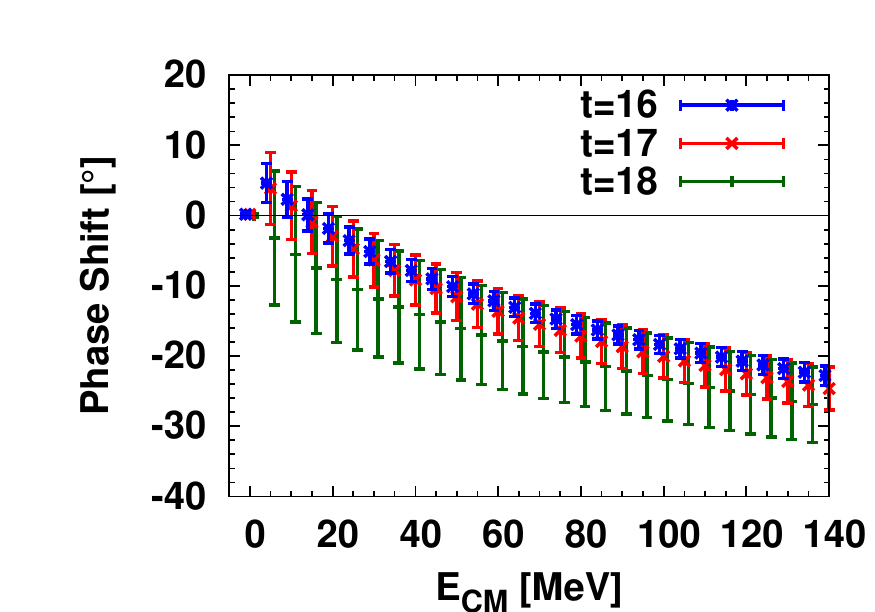}
%
%\vspace*{-2mm}
\caption{
\label{fig:phase:XiXi:3S1:eff}
$\Xi\Xi$ phase shifts in effective $^3S_1$ $(I=0)$ channel
obtained at $t = 16-18$.
}
\end{center}
%\vspace*{-3mm}
\end{figure}
%%%%%%%%%%%%%%%%%%%%%%%%%%%%%%%%%%%%%%%%%%%%%%%%%%

We first present the results for the $\Xi\Xi$ system in $^1S_0$ (iso-triplet) channel.
This channel belongs to the 27-plet %representation 
in flavor SU(3) classification
as does the $NN (^1S_0)$ system.
%and thus $\Xi\Xi (^1S_0) $serves as a ``doorway'' to probe the $NN (^1S_0)$ interaction,
%where the signal in the former is much cleaner than the latter on a lattice.
%In addition,
Since the ``dineutron'' ($NN (^1S_0)$) is nearly bound due to the strong attraction,
it has been a long standing question whether or not
the 27-plet interaction with the SU(3) breaking effects
forms a bound $\Xi\Xi (^1S_0)$ state~\cite{Haidenbauer:2014rna}.
%~\cite{Haidenbauer:2014rna, Rijken:hyp2015}.

Shown in Fig.~\ref{fig:pot:XiXi:1S0:cen}
is the LQCD result
for the central force $V_C(r)$
in the $\Xi\Xi (^1S_0)$ channel.
We observe a clear signal of 
the mid- and long-range attraction as well as the repulsive core at short-range,
resembling the phenomenological potential in $NN(^1S_0)$ system.
Within statistical fluctuations,
the results are found to be consistent with each other in the range $t = 16-18$,
which suggests that the contaminations from inelastic excited states are suppressed
and higher-order terms in the velocity expansion are small.
In Fig.~\ref{fig:phase:XiXi:1S0:cen},
we show the
%corresponding
phase shifts in terms of the center-of-mass energy.
It is found that the interaction is strongly attractive at low energies,
while it is not sufficient to form a bound $\Xi\Xi (^1S_0)$ state.
%The strong attraction observed in lattice 
%may be probed, e.g., as strong $\Xi\Xi$ correlation in heavy-ion collision experiments.
Such a strong attraction may be observed by, e.g., heavy-ion collision experiments.

We next consider the $\Xi\Xi$ system in $^3S_1$-$^3D_1$ (iso-singlet) coupled channel.
This channel belongs to the 10-plet in flavor SU(3),
a unique representation with hyperon degrees of freedom.
By solving the coupled channel Schr\"odinger equation, % with NBS correlators,
the central and tensor forces are obtained,
which are shown in left and right panels in Fig.~\ref{fig:pot:XiXi:3S1:cen_ten}, respectively.
The strong repulsive core is observed in the central force,
which 
%is in accordance with 
can be understood from the viewpoint of
the quark Pauli blocking effect~\cite{Aoki:2012tk, Oka:1986fr}.
There also exists an indication of a weak attraction at mid range which
may reflect the effect of small attractive one-pion exchange potential (OPEP).
The $\Xi\Xi$ tensor force is found to
be weaker with an opposite sign
compared to 
the $NN$ tensor force (right panel of Fig.~\ref{fig:pot:NN:3S1:cen_ten}).
This could be understood by the phenomenological 
one-boson exchange potentials,
where $\eta$ gives weaker and positive tensor forces
and $\pi$ gives much weaker and negative tensor forces.
%with flavor SU(3) meson-baryon couplings
%together with $F/D$ ratio by SU(6) quark model.
%
%
We also calculate the effective central potential in $^3S_1$ channel
by solving the (S-wave) single channel Schr\"odinger equation.
The effects of the tensor force are implicitly included
as $^3S_1$ $\rightarrow$ $^3D_1$ $\rightarrow$ $^3S_1$.
The corresponding phase shifts in effective $^3S_1$ channel are
given in Fig.~\ref{fig:phase:XiXi:3S1:eff},
which reflect the repulsive nature of the interaction.

%%%%%%%%%%%%%%%%%%%%%%%%%%%%%%%%%%%%%%%%%%%%%%%%%%%%
%\vspace*{-1mm}
\section{$NN$ systems ($S=0$ channel)}
%\vspace*{-1mm}
\label{sec:NN}

%%%%%%%%%%%%%%%%%%%%%%%%%%%%%%%%%%%%%%%%%%%%%%%%%%
\begin{figure}[t]
\begin{center}
%
%\vspace*{-4mm}
\includegraphics[angle=0,width=0.432\textwidth]{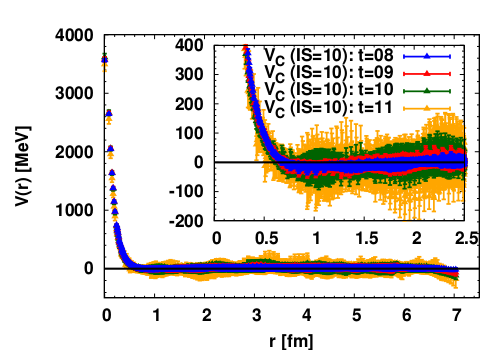}
%
%\vspace*{-2mm}
\caption{
\label{fig:pot:NN:1S0:cen}
$NN$ central force $V_C(r)$ in $^1S_0$ $(I=1)$ channel
obtained at $t = 8-11$.
}
\end{center}
%\vspace*{-3mm}
\end{figure}
%%%%%%%%%%%%%%%%%%%%%%%%%%%%%%%%%%%%%%%%%%%%%%%%%%

We present the results for the $NN$ system in $^1S_0$ (iso-triplet) channel.
In Fig.~\ref{fig:pot:NN:1S0:cen},
we show the central force $V_C(r)$ obtained at $t= 8-11$.
While the results suffer from large statistical fluctuations,
the repulsive core at short-range is clearly observed.
Its magnitude is more enhanced compared to that in $\Xi\Xi(^1S_0)$,
which can be understood from the one-gluon exchange picture.
We also observe that the potential is attractive at mid- and long-range,
resembling the phenomenological potential as OPEP.
In order to suppress the contaminations from inelastic states,
it is desirable to take larger $t$,
while the statistical fluctuations become larger.
Further studies with larger statistics are currently underway.

%%%%%%%%%%%%%%%%%%%%%%%%%%%%%%%%%%%%%%%%%%%%%%%%%%
\begin{figure}[t]
\begin{minipage}{0.48\textwidth}
\begin{center}
%
%\vspace*{-4mm}
\includegraphics[angle=0,width=0.90\textwidth]{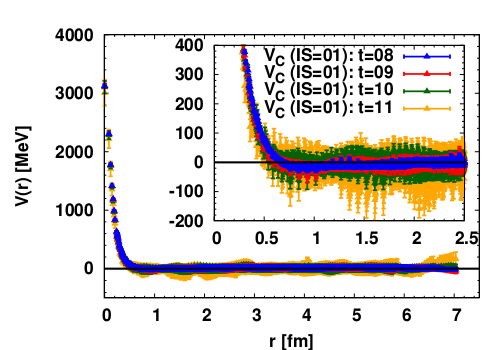}
%
%\vspace*{-2mm}
%\caption{
%}
%
\end{center}
\end{minipage}
\hfill
%\vspace*{-2mm}
\begin{minipage}{0.48\textwidth}
\begin{center}
%
%\vspace*{-4mm}
\includegraphics[angle=0,width=0.90\textwidth]{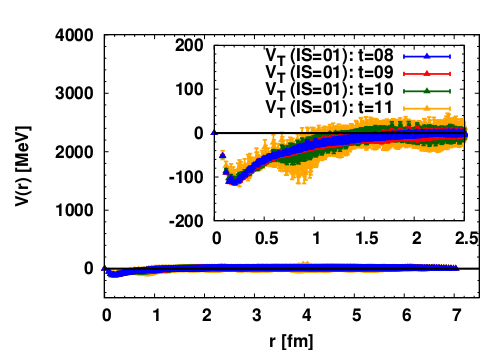}
%
%\vspace*{-2mm}
%\caption{
%}
%
\end{center}
\end{minipage}
\caption{
\label{fig:pot:NN:3S1:cen_ten}
$NN$ central force $V_C(r)$ (left) and tensor force $V_T(r)$ (right) in $^3S_1$-$^3D_1$ $(I=0)$ channel
obtained at $t = 8-11$.
}
%\vspace*{-3mm}
\end{figure}
%%%%%%%%%%%%%%%%%%%%%%%%%%%%%%%%%%%%%%%%%%%%%%%%%%

We next consider the $^3S_1$-$^3D_1$ (iso-singlet) channel.
In the left and right panels of Fig.~\ref{fig:pot:NN:3S1:cen_ten}, we show the
central force $V_C(r)$ and tensor force $V_T(r)$, respectively.
The most notable observation is 
the strong tensor force with the long-range tail structure.
Compared to the lattice tensor forces obtained with heavier quark masses~\cite{Aoki:2012tk},
the range of interaction is found to be longer.
In the central force, 
the repulsive core at short-range
as well as mid- and long-range attraction are observed.
Compared to the results of $NN$ $(^1S_0)$ channel,
there is a tendency that the interaction is more attractive.
Note that the results at larger $t$ are more reliable
since the contaminations from inelastic states are more suppressed.
%while the statistical fluctuations are larger.
The calculation of phase shifts with larger statistics is in progress.
In comparison to the empirical phase shifts,
it is also interesting to study the effect of
the heavier pion mass in this simulation ($m_\pi \simeq 146$ MeV).

%%%%%%%%%%%%%%%%%%%%%%%%%%%%%%%%%%%%%%%%%%%%%%%%%%%%
%\vspace*{-1mm}
\section{$N\Omega$ system ($S=-3$ channel)}
%\vspace*{-1mm}
\label{sec:NOmg}

\begin{figure}[t]
\begin{minipage}{0.48\textwidth}
\begin{center}
%
%\vspace*{-4mm}
\includegraphics[angle=0,width=0.8\textwidth]{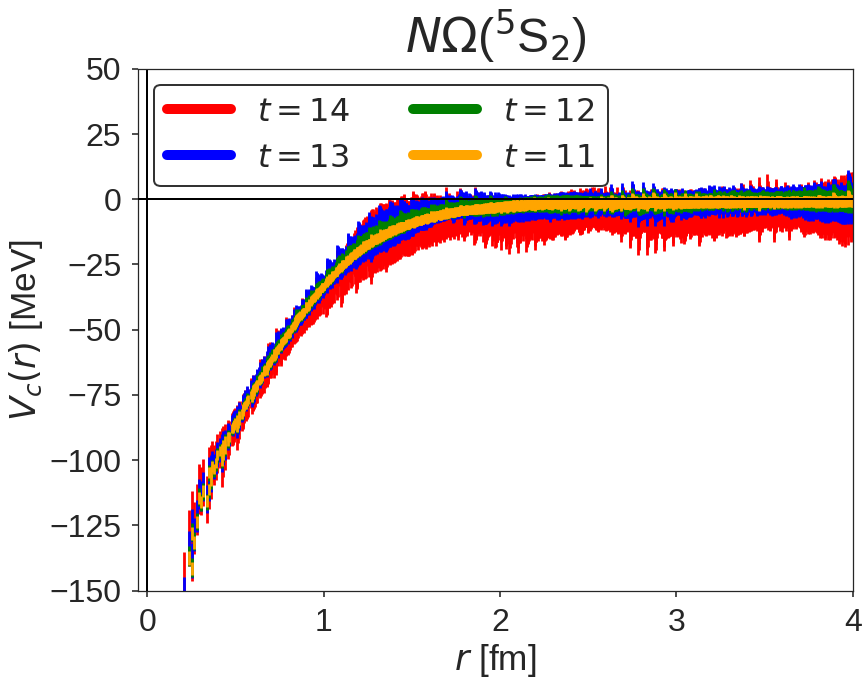}
%
%\vspace*{-2mm}
\caption{
\label{fig:pot:NOmg:5S2:cen}
$N\Omega$ central force $V_C(r)$ in $^5S_2$ channel
obtained at $t = 11-14$.
}
\end{center}
\end{minipage}
\hfill
%\vspace*{-2mm}
\begin{minipage}{0.48\textwidth}
\begin{center}
%
%\vspace*{-4mm}
\includegraphics[angle=0,width=0.8\textwidth]{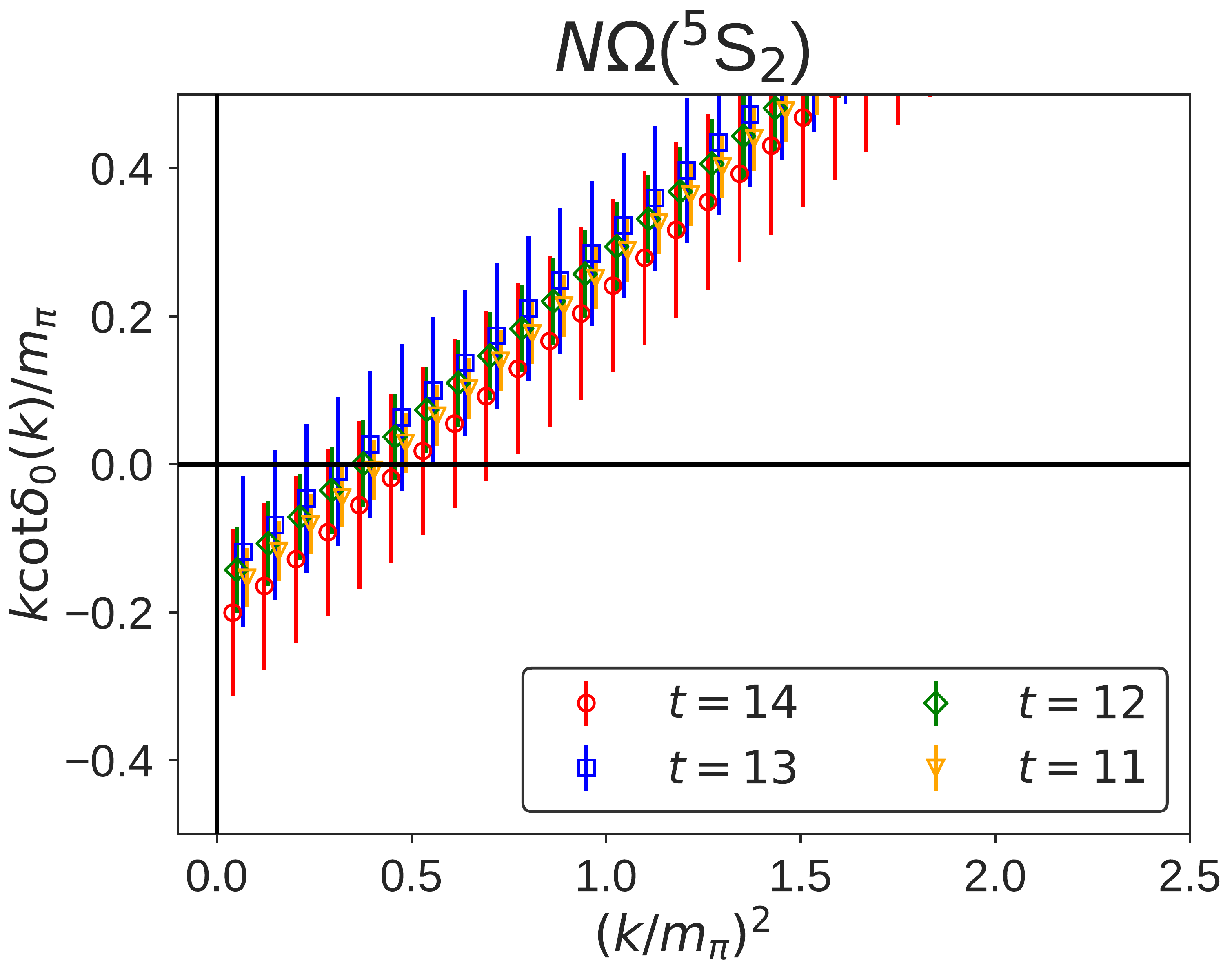}
%
%\vspace*{-2mm}
\caption{
\label{fig:phase:NOmg:5S2:cen}
$k\cot\delta_0(k)/m_\pi$ in $N\Omega (^5S_2)$ channel
obtained at $t = 11-14$.
}
\end{center}
\end{minipage}
%\vspace*{-3mm}
\end{figure}

We show the results of $N\Omega$ interaction in $^5S_2$ channel.
Shown in Fig.~\ref{fig:pot:NOmg:5S2:cen}
is the central force $V_C(r)$ obtained at $t = 11-14$.
It is found that the interaction is attractive at all distances.
Note that no Pauli blocking effect is present between $N$ and $\Omega$.
The potential is found to be stable against the change of $t$, which implies that
the coupling to the two octet baryon systems are suppressed
and
higher-order terms in the velocity expansion are small.
We calculate the corresponding phase shifts $\delta_0(k)$ 
and present them in the form of $k \cot\delta_0(k)/m_\pi$ as a function of $(k/m_\pi)^2$
in Fig.~\ref{fig:phase:NOmg:5S2:cen}.
The data are aligned almost linearly as is expected by the
effective range expansion at low energies, where
the intercept at $k^2=0$ corresponds to the inverse of the scattering length
and the slope corresponds to the (half of) the effective range.
While the statistical fluctuations are still large,
the results imply that the scattering length is negative (in particle physics convention)
and the system is bound (compared to the $N\Omega$ threshold).
Such a strong attraction can be studied through the
$N\Omega$ pair-momentum correlation in heavy-ion collision experiments~\cite{Morita:2016auo}.

%%%%%%%%%%%%%%%%%%%%%%%%%%%%%%%%%%%%%%%%%%%%%%%%%%
%%%%%%%%%%%%%%%%%%%%%%%%%%%%%%%%%%%%%%%%%%%%%%%%%%

%%%%%%%%%%%%%%%%%%%%%%%%%%%%%%%%%%%%%%%%%%%%%%%%%%
%\vspace*{-1mm}
\section{Summary}
%\vspace*{-1mm}
\label{sec:summary}

We have presented the latest lattice QCD results
for baryon interactions
at nearly physical quark masses,
$m_\pi \simeq 146$ MeV and $m_K \simeq 525$ MeV
on a large lattice box $(96 a)^4 \simeq (8.1 {\rm fm})^4$.
Baryon forces have been calculated from 
%Nambu-Bethe-Salpeter (NBS) 
NBS 
correlators
in the (time-dependent) HAL QCD method.
We have shown preliminary results for $\Xi\Xi$, $NN$ and $N\Omega$ interactions.

In the $\Xi\Xi (^1S_0)$ channel, a strong attraction is obtained,
although it is not strong enough to form a bound state.
In the $\Xi\Xi$ ($^3S_1$-$^3D_1$) channel,
the strong repulsive core exists in the central force.
Tensor force is found to be weak and have an opposite sign compared to the $NN$ tensor force.
For $NN$ forces,
the strong tensor force is clearly obtained.
Repulsive cores as well as mid- and long-range attractions have been observed 
in central forces in both $^1S_0$ and $^3S_1$-$^3D_1$ channels,
where repulsive core in $NN(^1S_0)$ is stronger than $\Xi\Xi(^1S_0)$.
In the $N\Omega (^5S_2)$ channel,
the interaction is found to be attractive at all distances
and a bound state (compared to the $N\Omega$ threshold) is possibly formed.
Strong attractions observed in $\Xi\Xi (^1S_0)$ and $N\Omega (^5S_2)$
can be best searched through pair-momentum correlations in heavy-ion collision experiments.
The physical interpretations for these baryon interactions
from the point of view of
quark Pauli blocking effect,
one-gluon exchange at short distance
and
one-boson exchange potentials at mid to long distance
are discussed.

The analysis with increased statistics is in progress.
In future, it is also important to study baryon interactions in parity-odd channel
including spin-orbit forces~\cite{Murano:2013xxa} and three-baryon forces~\cite{Doi:2011gq},
all of which have a large impact to determine the EoS of dense matter based on lattice QCD~\cite{Inoue:2013nfe}.

%%%%%%%%%%%%%%%%%%%%%%%%%%%%%%%%%%%%%%%%%%%%%%%%%%%%
%\vspace*{-1mm}
\section*{Acknowledgments}
%\vspace*{-1mm}

%We thank ... for collaborations.
We thank members of PACS Collaboration for the gauge configuration generation.
The lattice QCD calculations have been performed on the K computer at RIKEN, AICS
(hp120281, hp130023, hp140209, hp150223, hp150262, hp160211, hp170230),
HOKUSAI FX100 computer at RIKEN, Wako (G15023, G16030, G17002)
and HA-PACS at University of Tsukuba (14a-20, 15a-30).
T.D. and T.H. are supported in part by RIKEN iTHES Project and iTHEMS Program.
We thank ILDG/JLDG~\cite{conf:ildg/jldg}
which serves as an essential infrastructure in this study.
This work is supported in part by 
MEXT Grant-in-Aid for Scientific Research (JP15K17667),
SPIRE (Strategic Program for Innovative REsearch) Field 5 project,
"Priority Issue on Post-K computer" (Elucidation of the Fundamental Laws and Evolution of the Universe)
and
Joint Institute for Computational Fundamental Science (JICFuS).

%%%%%%%%%%%%%%%%%%%%%%%%%%%%%%%%%%%%%%%%%%%%%%%%%%%%
\clearpage
%\vspace*{-3mm}
%\bibliography{lattice2017}

%%%%%%%%%%%%%%%%%%%%%%%%%%%%%%%%%%%%%%%%%%%%%%%%%%%%%%%%%%%%%%%%%%%%%%%%%%%%%
\end{document}